\pdfoutput=1
\documentclass[prc,preprint,eqsecnum,nofootinbib,amsmath,amssymb,preprintnumbers,tightenlines,longbibliography,aps]{revtex4-1}
\usepackage{afterpage}
\usepackage{mathrsfs}
\usepackage{graphicx}% Include figure files
\usepackage{bm}% bold math
\usepackage{paralist}
\usepackage[colorlinks,linkcolor=blue,citecolor=blue]{hyperref}

\def\e{{\epsilon} }

\def\ls2{{\ell_s^2}}

\newcommand\bga{\begin{align}}
\newcommand\nda{\end{align}}

\def\E{{\bf E}}

\def\st{\begin{equation}}
\def\stp{\end{equation}}
\def\bg{\begin{eqnarray}}
\def\nd{\end{eqnarray}}
\def\Eq#1{Eq.~(\ref{#1})}
\def\Eqs#1{Eqs.~(\ref{#1})}
\def\App#1{Appendix~\ref{#1}}
\def\Fig#1{Fig.~\ref{#1}}
\def\Sect#1{Section~\ref{#1}}
\def\Ref#1{Ref.~\cite{#1}}

\def\e{\varepsilon}

\def\nott#1{\setbox0=\hbox{$#1$}                % set a box for #1 
   \dimen0=\wd0                                 % and get its size
   \setbox1=\hbox{/} \dimen1=\wd1               % get size of /
   \ifdim\dimen0>\dimen1                        % #1 is bigger
      \rlap{\hbox to \dimen0{\hfil/\hfil}}      % so center / in box
      #1                                        % and print #1
   \else                                        % / is bigger
      \rlap{\hbox to \dimen1{\hfil$#1$\hfil}}   % so center #1
      /                                         % and print /
   \fi}                                         %

\advance\parskip 1.9pt
\advance\voffset -0.2in

\def\bra{\langle}
\def\ket{\rangle}

\def\E{\mathcal{E}}
\def\P{\mathcal{P}}
\def\d{\delta}
\def\dz{\delta_{z}}

\def\be{\begin{equation}}
\def\ee{\end{equation}}
\def\bea{\begin{eqnarray}}
\def\eea{\end{eqnarray}}

\begin{document}

\title{Plane correlations in small colliding systems}

\author{Li Yan}
\email{li.yan@cea.fr}
\affiliation
    {%
	CNRS, URA2306, IPhT, 
	Institut de Physique Th\'eorique de Saclay, F-91191
	Gif-sur-Yvette, France \\
    }%

\begin{abstract}

%Geometric effects on 
%correlations of flow harmonics are studied for small collision
%systems: p-Pb, d-Au and $^3$He-Au. 
%Strong anti-correlation between ellipticity and triangularity is discovered
%from Monte Carlo Glauber simulations for the small collision systems, 
%due to fluctuations and an intrinsic background geometry.
%Similar but even stronger
%correlation or anti-correlation patterns are found when dipolar asymmetry is involved. 
%Correlations between flow harmonics 
%are accordingly expected due to medium collective expansion.  
%Especially, correlation between elliptic flow and triangular flow
%is proposed as a probe for the test
%of medium collectivity in central p-Pb collisions.
 
I propose event-plane correlations as a test of collectivity 
in small colliding systems:
In d-Au and $^3$He-Au collisions, I predict a strong anti-correlation 
between elliptic and triangular flow, generated by the geometry of 
the light projectile. A significant anti-correlation is also predicted 
in central p-Pb collisions at the LHC, which is solely generated by fluctuations.
Similar, but stronger correlation patterns are predicted 
in correlations involving dipolar flow.

\end{abstract}

\maketitle

\section{Introduction}

In high energy nucleus-nucleus collisions, collectivity is best manifested 
in the measured harmonic flow~\cite{Heinz:2013th,Luzum:2013yya}.
Defined as the 
Fourier harmonics of a single-particle spectrum,
\be
\label{vn_def}
\frac{dN}{d\phi_p} = \frac{N}{2\pi}[1+\sum_{n=1}^\infty V_n e^{-in\phi_p} + \text{complex conj.}]\,,
\ee
harmonic flow $V_n=v_n\exp(in\Psi_n)$ characterises final state anisotropy in the 
momentum space.
Recently, analyses of harmonic flow were extended to small colliding systems, regarding the 
striking observation of a ridge structure in long-range (in rapidity) 
two-particle correlations in high
multiplicity events~\cite{CMS:2012qk}. 
At RHIC energy, substantial elliptic flow $v_2$ 
and triangular flow $v_3$ were 
extracted in d-Au and $^3$He-Au collisions~\cite{Adare:2013piz,greenspan1996}. 
At the LHC, in $\sqrt{s_{\mbox{\tiny NN}}}=5.07$ TeV p-Pb collisions, 
it was also shown that harmonic flow can be as 
large as that measured in Pb-Pb collisions 
with $\sqrt{s_{\mbox{\tiny NN}}}=2.76$ TeV~\cite{Chatrchyan:2013nka,Aad:2014lta,ABELEV:2013wsa}, 
provided that multiplicity 
yields in these events are comparable~\cite{Basar:2013hea}. The 
observation of harmonic flow 
implies a possible collective expansion stage during the collisions of p-Pb, d-Au and $^3$He-Au,
which is further 
supported by various theoretical simulations.
With appropriate descriptions of the initial density profile, 
reasonable predictions on harmonic flow have been achieved 
for p-Pb~\cite{Bozek:2011if,Kozlov:2014fqa,Qin:2013bha}, 
d-Au~\cite{Qin:2013bha,Nagle:2013lja} and $^3$He-Au~\cite{Nagle:2013lja,Bozek:2014cya} 
colliding systems with viscous hydrodynamics. 
In a transport approach, it is also found that
the prediction of $v_2$ and $v_3$ requires sufficient interactions
in the late stage~\cite{Bzdak:2014dia}. 
 
However, the statement of collectivity in small 
colliding systems is still under debate.
In particular, it is challenged by the fact that 
long-range correlations seen in the experimental data can
originate alternatively from pre-equilibrium physics~\cite{Gyulassy:2014cfa}. %which gives rise to $v_2$ and $v_3$. 
A  way to disentangle competing models, and 
especially to clarify the issue of collectivity, lies in the 
measurements %of harmonic flow with higher precisions, 
%namely, determination 
of flow fluctuations and mixing between harmonics. 
The idea relies on the knowledge that medium collectivity %expansion
associates harmonic flow with %initial state geometrical fluctuations.
event-by-event fluctuations in the initial density profile.
The CMS Collaboration has 
updated its measurements of $v_2$ 
fluctuations in terms of cumulants $v_2\{m\}$ from 
multi-particle correlations~\cite{Wang:2014rja,Khachatryan:2015waa}.    
The observed pattern of $v_2\{m\}$ agrees with cumulants of 
initial ellipticity which is purely driven
by fluctuations~\cite{Yan:2013laa,Bzdak:2013rya}, 
which coincides with the expectation from collective expansion.

This work is motivated in a similar manner, but with emphasis
on the mixing of harmonic flow. 
Especially, %geometric effects of initial state 
event-by-event fluctuations in the initial density profile
dominate the
mixing between lower harmonics due to a linear eccentricity scaling
~\cite{Aad:2014fla,Teaney:2013dta}. 
%of the medium response.
This allows one to estimate the mixing between $V_2$ and $V_3$ %will be 
for p-Pb, d-Au and $^3$He-Au, 
in terms of
the mixing between initial ellipticity and triangularity.   
Analyses in this work are focused on central collision events,  
where the initial state 
geometries differ dramatically in p-Pb, 
d-Au and $^3$He-Au, with the background density
profile azimuthally symmetric,
dumbbell-shaped and triangle-shaped respectively.
In addition, correlations of harmonics involving a 
dipolar flow $V_1$ are studied as well,
based on initial mixing involving dipolar asymmetry.

In order to have a consistent comparison with experimental measurements, a series of
correlation coefficients are %used in this paper, which are defined 
specified in \Sect{sec:defs}.
Collision events are simulated by PHOBOS Monte Carlo Glauber model, with
details of the model simulation described in \Sect{sec:sim}. 
Correlations of initial anisotropies from simulations are presented in \Sect{sec:sim}.
%In \Sect{sec:res}, correlations involving $\E_1$ (or $V_1$) are also shown as predictions.
In \Sect{sec:dis}, analytical analysis of the correlations is
%carried out 
given based on an independent source 
approach, where initial state of the small colliding systems are modelled
by a number of independent sources.
Effects beyond linear eccentricity scaling are discussed with respect to
the mixing between $V_2$ and $V_3$ in p-Pb collisions in \Sect{sec:p23vv}.

\section{Correlation coefficients}
\label{sec:defs}
In this paper, to quantify
correlations between $V_n$ and $V_m$ (i.e., phase mixing between
$\Psi_n$ and $\Psi_m$ when $n\ne m$),
we define the following correlation coefficient which has a similar structure
as a Pearson correlation coefficient, 
\be
\label{pers_ini}
\P^{vv}_{nm}=\frac{Re\bra V_n^a(V_m^*)^b\ket}{\bra|V_n|^{2a}\ket^{1/2}\bra|V_m|^{2b} \ket^{1/2}}\,,
\ee
where $a$ and $b$ are integer numbers, satisfying $ma-nb=0$
due to a rotational symmetry. 
In \Eq{pers_ini} and throughout this paper, angular brackets 
$\bra\ldots\ket$ are used to notate average over events.
By definition one has 
$0\le|\P_{nm}^{vv}|\le1$, with $1$ stands for an absolute 
correlation or anti-correlation and $0$ corresponds to pure randomness.

$\P_{nm}^{vv}$ has been studied through the measurements of event-plane 
correlations in Pb-Pb systems by the ATLAS collaboration~\cite{Aad:2014fla},
using the scalar-product method.
In particular, the correlation between $V_2$ and $V_3$, 
\be
\label{eq:p23vv}
\P^{vv}_{23}=\frac{Re\bra V_2^3(V_3^*)^2\ket}{\bra|V_2|^{6}\ket^{1/2}\bra|V_3|^{4} \ket^{1/2}}\,,
\ee
was found negligibly weak for most of the centralities, 
except peripheral collisions. Unlike correlations involving 
higher harmonics, which receive significant contributions from non-linear 
flow response during medium expansion~\cite{Teaney:2013dta}, 
$\P^{vv}_{23}$ is mostly determined by the mixing between initial anisotropies.
Anisotropies of initial state are quantified generally by eccentricities.
In each single collision event, if one denotes 
average over transverse plane according to the initial density profile  
%according to, for instance, 
%initial energy density $\rho(r,\phi_r, \tau_o)$, 
as $\{\ldots\}$, the first three eccentricities are
\footnote{
Re-centering corrections are implied in these definitions, with $\{re^{i\phi_r}\}=0$
}
\begin{align}
\label{ep1}
\E_1=&\varepsilon_1e^{i\Phi_1}=-\frac{\{r^3 e^{i\phi_r}\}}{\{r^3\}}\,,\\
\label{ep2}
\E_2=&\varepsilon_2e^{i2\Phi_2}=-\frac{\{r^2 e^{2i\phi_r}\}}{\{r^2\}}\,,\\
\label{ep3}
\E_3=&\varepsilon_3e^{i3\Phi_3}=-\frac{\{r^3 e^{3i\phi_r}\}}{\{r^3\}}\,.
\end{align} 
In accordance with \Eq{vn_def}, complex notations have been applied in
\Eqs{ep1}, (\ref{ep2}) and (\ref{ep3}), so that both magnitude and 
phase of eccentricity are defined simultaneously. 
Approximately a linear eccentricity scaling can be assumed for lower harmonics,
based on event-by-event hydrodynamic simulations~\cite{Niemi:2012aj},
\be
\label{eq:lresp}
V_n = \kappa_n\E_n\,, \qquad n\le 3\,.
\ee
$\kappa_n$ in \Eq{eq:lresp} is known as the medium response coefficient,
which is determined by the property of medium collectivity.
It is then obvious that correlations of final state between harmonics
are identical to the corresponding correlations of initial anisotropies, 
e.g.,
\be
\label{eq:vvee}
\P_{23}^{vv}=\P_{23}^{\varepsilon\varepsilon}\,.
\ee
Note that the superscript with $\e$'s on the right hand side of \Eq{eq:vvee}
implies correlations of initial anisotropies.

%Although the weak
%mixing between $\E_2$ and $\E_3$ (or equivalently $V_2$ and $V_3$) is less interesting for 
%nucleus-nucleus collisions, it is expected to be a sensitive probe for small collision systems. 
Mixing between $\E_2$ and $\E_3$ (or equivalently $V_2$ and $V_3$) in small colliding
systems is of particular interest for two reasons. First, geometrical fluctuations
in the initial state are much more pronounced in p-Pb, d-Au and $^3$He-Au
than those in nucleus-nucleus collisions. 
For instance, the multiplicity in central p-Pb is 
smaller roughly by a factor of 10 than in central Pb-Pb collisions, one 
expects initial state fluctuations to be larger by a factor of $\sqrt{10}$. 
Second, background geometries in d-Au and $^3$He-Au are generically 
deformed. For events in the 
most central collision bins of p-Pb, d-Au and $^3$He-Au, %namely, events with highest multiplicty, 
geometry of the system is dominated by
the configuration of proton, deuteron and $^3$He respectively. Therefore,  
initial state background geometry of p-Pb is azimuthally symmetric.
But for d-Au and $^3$He-Au, a dumbbell-shaped initial density 
and an intrinsic triangle-shaped initial density 
are expected respectively.
%as displayed in \Fig{fig:dens}. 
Accordingly, non-trivial geometries lead to an intrinsic 
initial ellipticity $\bar{\E}_2$ in d-Au and 
an intrinsic initial triangularity $\bar{\E}_3$ in $^3$He-Au. 
Both fluctuations and background geometry will be shown
essential for the generation
of anisotropy correlations, later in \Sect{sec:sim} and \Sect{sec:dis}.
Similarly, one also expects fluctuations and geometric shape induce 
correlations involving a dipolar asymmetry, namely correlations of harmonics
involving $V_1$.
In this paper the following types of correlations are studied as well, 
\be
\P_{12}^{vv}, \quad \P_{13}^{vv} \quad \mbox{and} \quad\P_{123}^{vvv}\,.
\ee 
Note that $\P_{123}^{vvv}$ is defined as 
\be
\P_{123}^{vvv}=\frac{Re\bra V_1V_2V_3^*\ket}{\bra|V_1V_2|^{2}\ket^{1/2}\bra|V_3|^{2} \ket^{1/2}}\,,
\ee
which is normalized differently from the three-plane 
event-plane correlations introduced by the
ATLAS collaboration~\cite{Aad:2014fla} with the 
scalar-product method.

\section{Monte Carlo Glauber simulations}
\label{sec:sim}

\begin{figure}
\begin{center}
\includegraphics[width=0.8\textwidth]{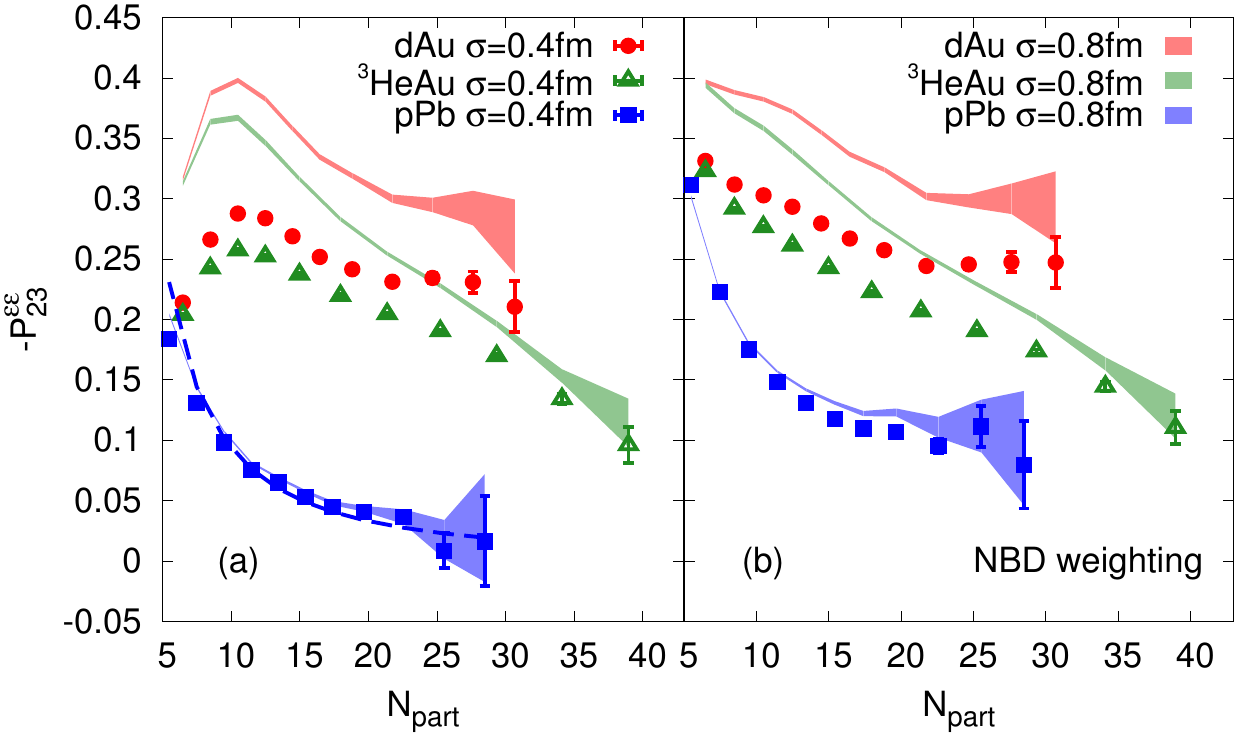}
\caption{\label{fig:p23ee}
(Color online)
Initial state mixing between $\E_2$ and $\E_3$ as a function of number of participants.
Results in panel (a) are from PHOBOS Monte Carlo Glauber model with a constant weighting of
$C_s$ in \Eq{entropy}. Dashed line is analytical expectation from the independent-source model. 
Results with an extra NBD weighting is shown in Panel (b). Results from smearing with $\sigma=0.4$ fm 
are plotted with symbols, while shaded bands correspond to the case with $\sigma=0.8$ fm. 
}
\end{center}
\end{figure}

\begin{figure}
\begin{center}
\includegraphics[width=0.9\textwidth]{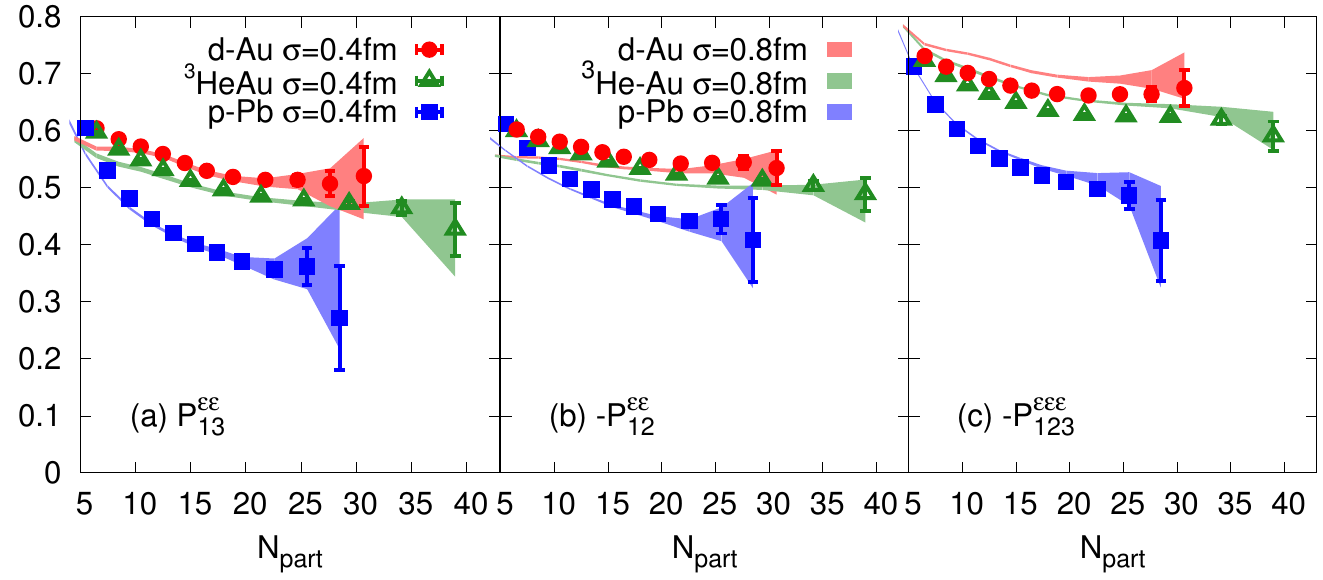}
\caption{\label{fig:pcall}
(Color online)
Initial state mixing involving $\E_1$ as a function of number of participants, 
represented by (a) $\P^{\varepsilon\varepsilon}_{12}$, 
(b) -$\P^{\varepsilon\varepsilon}_{13}$,
and (c) -$\P^{\varepsilon\varepsilon}_{123}$. All the results are 
obtained with an extra NBD weighting for the entropy 
production. Results with $\sigma=0.4$ fm are plotted with symbols,
while shaded bands correspond to $\sigma = 0.8$ fm.
}
\end{center}
\end{figure}

In this work, we simulate initial state of heavy-ion collisions 
using the PHOBOS Monte Carlo Glauber model~\cite{Alver:2008aq,Loizides:2014vua}. 
Additional modifications are taken into account with respect to small colliding
systems. For instance, the wounding profile of two colliding
nucleons is taken to be a Gaussian instead of a step function~\cite{Bozek:2013uha}.
Regarding heavy-ion collisions,
PHOBOS Glauber model captures the dominant geometric structure and effects of 
fluctuations of nucleons inside a colliding nucleus.     
On an event-by-event basis,
fluctuations of nucleon positioning inside nucleus are introduced 
randomly in model simulations, resulting in a 
discrete distribution of number of participants $N_{part}$ 
and collisions $N_{coll}$. 
%$N_{p}(x_i,y_i, \tau_o)$ and collisions $N_c(x_i,y_i,\tau_o)$. 
A Gaussian smearing centered at the location of each participant and collision is applied,  
\be
\frac{1}{2\pi \sigma^2}e^{-|\vec x_\perp|^2/2\sigma^2}\,.
\ee
The size of smearing is taken to be $\sigma=0.4$ fm and $0.8$ fm in this work for
analysis. Initial entropy density is thus obtained by assuming a 
proper weight associated with each participant and collision,
\be
\label{entropy}
s(x,y,\tau_o)=C_s\left[\frac{1-\alpha}{2}N_{part}(x,y) 
+ \alpha N_{coll}(x,y)\right]
\,.
\ee 
Instead of a constant weighting of $C_s$, 
a negative binomial distribution (NBD) weighting is known
necessary to reproduce the measured multiplicity 
distributions~\cite{Bozek:2013uha} in
small colliding systems, 
which introduces extra fluctuations. %as one should notice.
Especially, probability of generating high multiplicity events
is enhanced with a negative binomial weighting. 
In this work, following Refs.\cite{Bozek:2013uha,Qin:2013bha} 
and taking
the same set of parameters for p-Pb at the LHC, and d-Au and $^3$He-Au at RHIC,
ten million events are generated. 
Events are then classified with respect to number of participants $N_{part}$, 
which has a
correspondence to the centrality classification used in experiments.
For simulations with ten million events, the centrality of bin class 
with reliable statistics
can be as small as 0.01\%, which corresponds to the right most points in \Fig{fig:p23ee} and
\Fig{fig:pcall}.

Anti-correlations between $\E_2$ and $\E_3$ in p-Pb, d-Au and $^3$He-Au are 
discovered via model simulations, with results shown in \Fig{fig:p23ee}.
Note that a minus sign is implied.  
Panel (a) of \Fig{fig:p23ee} is obtained based on PHOBOS
Glauber model, without additional NBD weighting for the entropy production. Albeit less
realistic, the case with a constant weighting  
is closer to the model of independent sources~\cite{Bhalerao:2011bp,Bhalerao:2006tp,Blaizot:2014wba}, 
especially for the p-Pb system the correlation $\P_{23}^{\e\e}$ is found quantitatively
consistent with predictions from independent sources (see discussions around
\Eq{eq:p23ppb} in \Sect{sec:dis}).\footnote{
Note that in the original PHOBOS Glauber model\cite{Alver:2008aq,Loizides:2014vua}, 
a repulsive correlation is applied 
in the simulations, so it is not strictly equivalent to a model of 
independent sources.
}
Simulations with NBD weighting %described in \Sect{sec:sim}, 
which correspond to experiments 
carried out at RHIC and the LHC are presented in 
\Fig{fig:p23ee} (b). Errors of these results are estimated based on 
variance. Effect of smearing is investigated with $\sigma=0.4$ fm (symbols with
error bars) and $\sigma=0.8$ fm (shaded bands). Larger smearing size 
leads to stronger correlations between $\E_2$ and $\E_3$, 
which is more significant for d-Au and $^3$He-Au systems.

In all three colliding systems, 
the correlation between $\E_2$ and $\E_3$ decreases towards central
collisions. 
This decrease is associated with the decrease of fluctuations in the 
initial state. When extra fluctuations are introduced from a
NBD weighting of entropy production, seen as changes from \Fig{fig:p23ee} 
(a) to \Fig{fig:p23ee} (b), correlations are enhanced as well.
Especially for p-Pb, the correlation of the
most central events reaches 10\%, as shown in \Fig{fig:p23ee} (b).
In addition, from \Fig{fig:p23ee} there is 
an apparent hierarchy relation of correlations among the three systems, 
\be
\label{eq:hie}
|\P_{23}^{\varepsilon\varepsilon}(\mbox{p-Pb})|
<|\P_{23}^{\varepsilon\varepsilon}(^3\mbox{He-Au})|
<|\P_{23}^{\varepsilon\varepsilon}(\mbox{d-Au})|
\ee
which reflects the differences in the background geometry of each 
system. One may na\"ively understand this relation by noticing that 
$\E_2$ has one power higher than $\E_3$ in the original definition of $\P_{23}^{\e\e}$, which
makes a backgound ellipticitiy more important than a background 
triangularity. More detailed discussion of this hierarchy relation
are given in \Sect{sec:dis}.

\Fig{fig:pcall} presents initial correlations involving $\E_1$, as predictions for 
the flow correlations involving $V_1$. Again, to have all results on the same scale
a minus sign is applied when necessary. 
Irrespective of the detailed background geometry, in all the three systems  
a negative $\P_{12}^{\e\e}$ and a negative $\P_{123}^{\e\e\e}$ are obtained 
which indicate
anti-correlations between $\E_1$ and $\E_2$, and among $\E_1$, $\E_2$ and $\E_3$ 
respectively,
while $\E_1$ and $\E_3$ are found strongly correlated.
Although 
these correlations are stronger than $\P_{23}^{\e\e}$, they exhibit very similar 
dependence on the background geometry of the system and fluctuations, i.e., 
similar hierarchy relations regarding the three systems.
More detailed discussions are postponed to \Sect{sec:dis}.
It is also
worth mentioning that for the most central events in d-Au systems, 
correlations from simulations are compatible with those expected in peripheral 
nucleus-nucleus collisions~\cite{Bhalerao:2011yg,Teaney:2010vd,Jia:2012ma}, 
where the effects of fluctuations are comparable. 

Measuring event-plane correlations in small collision systems is a 
challenging analysis. In \App{app:sta}, we evaluate the order of magnitude of 
the statistical error. We find that in order to detect $\P_{23}^{vv}$ 
with a few percent accuracy in p-Pb central collisions at the LHC, one 
typically needs a million events per centrality bin.

\section{Understanding of initial state correlations}

\label{sec:dis}

\begin{figure}
\begin{center}
\includegraphics[width=0.7\textwidth]{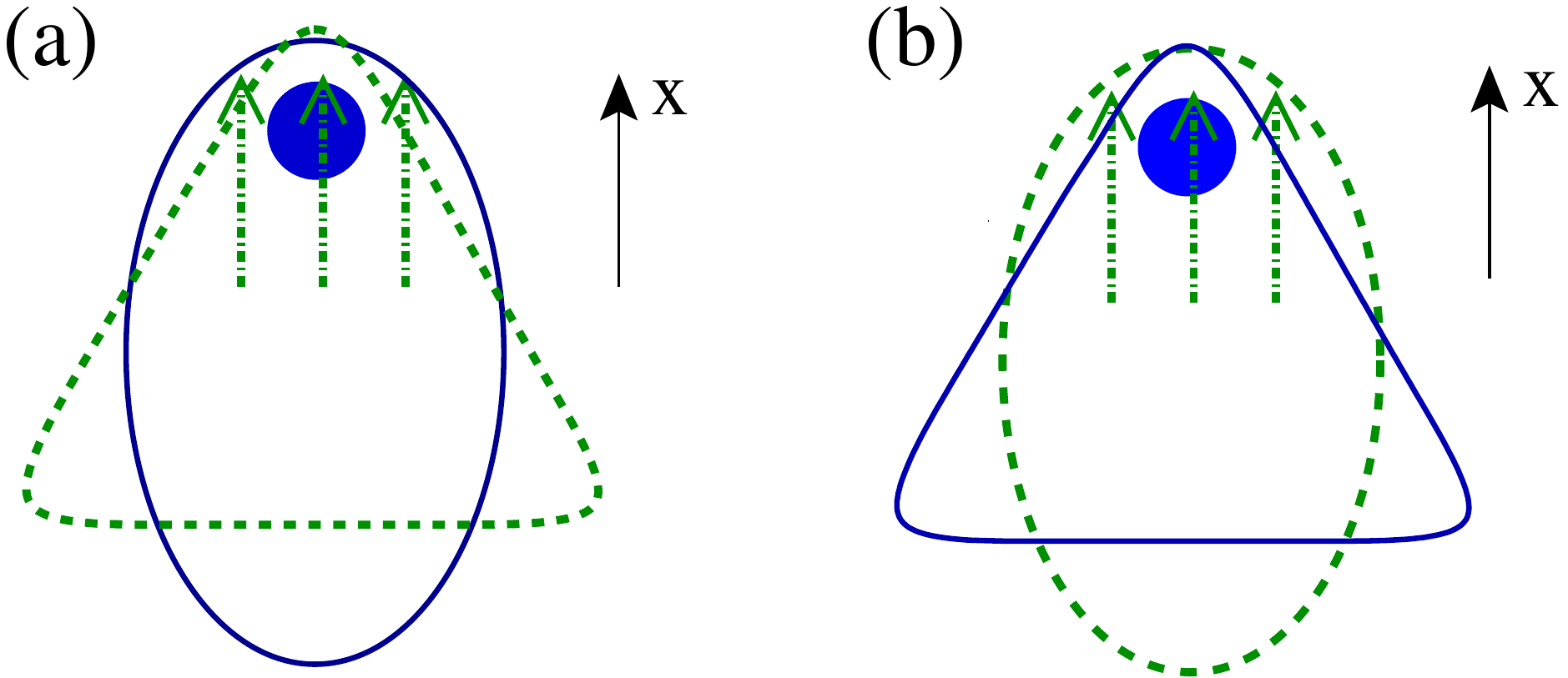}
\caption{\label{fig:toy1}
(Color online)
Alignment of anisotropies due to fluctuations on top
of smooth background (solid blue shape) with a net ellipticity $\E_2$ (a)
and a net triangularity $\E_3$ (b). Green dashed shapes
correspond to anisotropies generated by one addition
hot spot in the tip area of the background. Dipolar 
asymmetry is illustrated as green arrows.
}
\end{center}
\end{figure}

The origin of anisotropy correlations in the initial state
of heavy-ion collisions, with respect to the effects of 
fluctuations and a background geometry, 
can be understood qualitatively as follows.
In heavy-ion collisions, a bumpy initial state density profile 
can be equivalently realized by randomly throwing additional 
spots on top of a smooth background. 
For a smooth background with a net ellipticity or triangularity,
the area of tip is \emph{always larger} than that of side, 
which is purely a consequence of geometric effect.
Therefore, 
one expects the additional spots 
to appear more probably in the area of tip rather than
side.
\Fig{fig:toy1} describes anisotropies induced by
one additional hot spot in the tip area of an elliptic 
background (a), and a triangular background (b).
In both cases, one can check that the 
induced anisotropies are aligned with the background shape
in a specified way: $\Phi_1=\pi$, $\Phi_2=\pi/2$,
and $\Phi_3=\pi/3$,
which generates perfect \emph{anti-correlations} between $\E_2$ and $\E_3$,
$\E_1$ and $\E_2$, and among $\E_1$, $\E_2$ and $\E_3$, and perfect \emph{correlation}
between $\E_1$ and $\E_3$.
For a more sophisticated density profile which fluctuates from
event to event, correlations are weaken due to contributions from configurations 
with excessive density around sides of the background shape, 
but the sign of correlation should not be changed.

%In the pure geometrical picture, 
%one may even have a quantitative estimate of the strength of plane 
%correlations. On an event-by-event basis, the strength of correlations reflects
%relative probablity of the two cases mentioned above, namely, the relative
%area of the tip and the side of the background shape. Accounting for the
%dumbbell-shaped background in a d-Au system, in which the area of the tip is

%\subsection{Anisotropy correlations from independent sources}
To a quantitative level,
the effects of fluctuations and background geometry on
correlations can be
studied in the independent-source model~\cite{Bhalerao:2011bp},
in which an initial state density profile is modelled by N point-like 
independent sources on top of a specified background.
Consequently, fluctuations of any quantity $f$ is characterized as
\be
\delta_f = \{f\} - \bra f\ket\,,
\ee
where $\{\ldots\}$ and $\bra\ldots\ket$ are used in the similar way as in
\Eqs{ep1} and (\ref{pers_ini}), indicating average
in the transverse plane for one single event and average over
events respectively.
Independency of sources allows one to write event averaged quantities in
powers of $1/N$, i.e., fluctuations. For instance, the two-point function 
which quantifies event average of quadratic order of fluctuations, is of
the order of $1/N$,
\be
\label{eq:twop}
\bra\delta_f\delta_g\ket = \frac{\bra fg\ket-\bra f\ket\bra g\ket}{N}\,.
\ee
Event average of higher order of fluctuations leads to higher order 
dependence on $1/N$.
More details of the model can be found in \cite{Bhalerao:2011bp} 
and in \App{app:indep}.
For p-Pb collisions, background is azimuthally symmetric and thus 
$\bar{\E}_2=\bar{\E}_3=0$.
One accordingly finds that the dominant contribution is purely driven
by fluctuations, and of the order of $1/N^{3/2}$, 
\be
\label{eq:p23ppb}
\P_{23}^{\varepsilon\varepsilon}(\mbox{p-Pb})=-\frac{1}{N^{3/2}}
\frac{\bra r^{12}\ket+6\bra r^6\ket^2 -18\bra r^4\ket\bra r^8\ket}
{\sqrt{12}\bra r^4\ket^{3/2}\bra r^6\ket} + O(1/N^{5/2})\,.
\ee
For a Gaussian smearing, the geometric factor in \Eq{eq:p23ppb} 
is known explicitly ($\approx 1.22$) and thus
an analytical description can be obtained. As shown as the blue dashed 
line in \Fig{fig:p23ee} (a), the prediction from independent sources naturally captures
the correlation pattern if one assumes that the number of independent sources
scales like
$N_{part}\sim 0.55 N$. Note also that in \Eq{eq:p23ppb} the correlation
is insensitive to a Gaussian smearing, which is indeed confirmed in \Fig{fig:p23ee}
that smearing $\sigma$ has little influence on the simulated correlation   
$\P_{23}^{\varepsilon\varepsilon}(\mbox{p-Pb})$.

In principle, the generic correlations induced by the nuclear configurations of
deuteron and $^3$He prevent applying the independent-source model
to the collisions of d-Au and $^3$He-Au. Nevertheless, 
the effect of a non-trivial background geometry is worth analyzing, at
least qualitatively.
For $^3$He-Au collisions, the conditions $\bar{\E}_2=0$ and $\bar{\E}_3\ne0$
allow terms of lower order in $1/N$ contribute. While for d-Au systems,
a non-zero background ellipticity picks up even lower order terms in 
the expansion with respect to $1/N$. It can also be understood directly from the definition of 
$\P_{23}^{\e\e}\sim\bra \E_2^3(\E_3^*)^2\ket$, that $\E_2$ has one power higher 
than $\E_3$. In total, counting the extra suppression due to fluctuations,
one finds
\be
|\P_{23}^{\varepsilon\varepsilon}(^3\mbox{He-Au})|\sim O\left(\frac{1}{N^{1/2}}\right),
\qquad
|\P_{23}^{\varepsilon\varepsilon}(\mbox{d-Au})|\sim O(1)\,,
\ee
which is consistent with \Eq{eq:hie}.
It is also interesting to notice that the hierarchy relation in \Eq{eq:hie}
disappears when $N_{part}$
is sufficiently small, as depicted in \Fig{fig:p23ee}, where only one of the 
constituent nucleons in deuteron and $^3$He participates the collision. 

For correlations involving a dipolar asymmetry, 
an ideal assumption would be taking $\E_1$ 
as a result of fluctuations in all the three small systems,
then the similar argument leads to
\be
\label{eq:p123_hie}
|\P_{123}^{\e\e\e}(\mbox{d-Au})|\sim|\P_{123}^{\e\e\e}(^3\mbox{He-Au})|\sim O(1)\,,
\qquad
|\P_{123}^{\e\e\e}(\mbox{p-Pb})|\sim O\left(\frac{1}{\sqrt{N}}\right)\,,
\ee
due to the fact that $\E_2$ and $\E_3$ are of the same power in the definition of 
$\P_{123}^{\e\e\e}$.
And indeed \Eq{eq:p123_hie} agrees with the results depicted in \Fig{fig:pcall} (c).
However, when the role of $\E_1$ becomes more important, as in $\P_{12}^{\e\e}$
and $\P_{13}^{\e\e}$ which contain $\E_1^2$ and $\E_1^3$ respectively,
one has to take into account the effect of an intrinsic $\bar \E_1$ in both d-Au and $^3$He-Au
\footnote{Note that there is no intrinsic $\bar \E_1$ in p-Pb.}.
Because a larger intrinsic $\bar \E_1$ is expected with repect to a dumbbell-shape
than a triangular shape, the correlations $\P_{12}^{\e\e}$ and $\P_{13}^{\e\e}$
are stronger in d-Au than in $^3$He-Au.

%On the other hand, an intrinsic $\E_2$ plays an significant role
%in $\P_{12}^{\e\e}$, from which one would expect
%\be
%|\P_{12}^{\e\e}(\mbox{d-Au})|\sim O(1),\qquad
%|\P_{12}^{\e\e}(\mbox{p-Pb})|\sim |\P_{12}^{\e\e}(^3\mbox{He-Au})|\sim 
%O\left(\frac{1}{\sqrt{N}}\right)\,,
%\ee
%which is not consistent with our model simulations. 

\section{$\P_{23}^{vv}$ in proton-lead beyond linear eccentricity scaling}
\label{sec:p23vv}

One must go beyond the approximate linear eccentricity scaling
to have a more realistic estimate for the correlations of harmonics
in experiments. For the central p-Pb collisions, 
non-linear flow response can still be ignored since background geometry 
is azimuthally symmetric, % and thus $V_2$ is not extraordinarily large, 
which case is similar to the ultra-central Pb-Pb. The most significant
corrections come from fluctuations in hydro response.
%
%Phase mixing between harmonic flow receives extra contributions 
%from late stage evolution in heavy-ion collisions. Despite the 
%non-linear flow generation which dominates mixing between
%higher harmonics and $V_2$ (or $V_3$), the most significant
%effect on $\P_{23}^{vv}$ in central p-Pb is expected to be fluctuations 
%of flow response from event to event.
While detailed knowledge of event-by-event fluctuations
of flow response requires a systematic analysis based on, e.g.,
event-by-event hydrodynamic simulations, it can be quantified % estimated 
by introducing an extra noise term in \Eq{eq:lresp}~\cite{Yan:2014nsa,Gardim:2011xv}
\be
\label{eq:lresp-n}
V_n =\kappa_n \E_n + X_n\,,
\ee
where the noise $X_n$ is complex. 
I further model $X_n$ as a Gaussian noise, which is uncorrelated
with the initial eccentricity. This noise breaks linear eccentricity
scaling~\cite{Niemi:2012aj}. The magnitude of the effect can be 
quantified by the Pearson correlation between the flow and the 
initial anisotropy,
%
%One is further allowed to assume  
%$X_n$ as Gaussian noise and uncorrelated with initial eccentricity 
%$\E_n$, which leads to an analytical expectation of the destructive effect on
%the linear scaling. It is manifested by a Pearson correlation
%between $V_n$ and $\E_n$,
\be
\label{eq:vep}
\frac{\bra V_n\E_n^*\ket}{\bra |V_n|^2\ket^{1/2} \bra |\E_n|^2\ket^{1/2}}
=1 - \lambda_n + O(\lambda_n^2)\,,
\ee
where
\be
\label{eq:lambda}
\lambda_n = \frac{1}{2}\frac{\bra |X_n|^2\ket}{\kappa_n^2\bra |\E_n|^2\ket}\,,
\ee
is a small and positive quantity, characterizing
the relative magnitude between fluctuations and average flow response. 
%Event-by-event hydro has shown 
%that $\lambda_n$ is a small quantity when $n=2$, $3$ in nucleus-nucleus collisions. 
For the correlation between $V_2$ and $V_3$, substituting \Eq{eq:lresp-n} into 
\Eq{eq:p23vv} results in
\be
\label{eq:pv23-n}
\P_{23}^{vv}=\P_{23}^{\e\e}\left[1-
\frac{9\bra|\E_2^4|\ket\bra|\E_2|^2 \ket}{\bra|\E_2|^6 \ket}\lambda_2
-\frac{4\bra|\E_3|^2 \ket^2}{\bra|\E_3|^4 \ket}\lambda_3 + O(\lambda^2)\right]\,.
\ee
The factors before $\lambda_2$ and $\lambda_3$ in \Eq{eq:pv23-n}
are determined by fluctuations of $\E_2$ and $\E_3$ respectively.
In p-Pb collisions, it was proposed that fluctuations of 
initial state eccentricities 
follow the so-called power distribution~\cite{Yan:2013laa},
from which the ratios between moments of initial state 
eccentricity are known explicitly. In central collision events,
approximately one has 
$9\bra|\E_2^4|\ket\bra|\E_2|^2 \ket/\bra|\E_2|^6 \ket \approx 3$
and $4\bra|\E_3|^2 \ket^2/\bra|\E_3|^4 \ket \approx 2$.

An estimate of $\lambda_2$ and $\lambda_3$ in p-Pb collisions 
at the LHC can be made in terms of the 
breaking of two-particle correlations. Fluctuations in hydro response
result in a breaking in the two-particle correlation function, 
which is characterised by the ratio~\cite{Gardim:2012im},
\be
r_n(p_T^a,p_T^b)=\frac{V_{n\Delta}(p_T^a, p_T^b)}
{\sqrt{V_{n\Delta}(p_T^a, p_T^a)V_{n\Delta}(p_T^b, p_T^b)}}\,.
\ee
Taking into account noise in hydro response in \Eq{eq:lresp-n},
one obtains
\be
r_n(p_T^a, p_T^b)=1-\lambda_n(p_T^a)-\lambda_n(p_T^b) + O(\lambda^2).
\ee
For the central collision events in the available $p_T$ range,
$r_2(p_T^a,p_T^b)$ from event-by-event hydrodynamic simulations 
can reach as low as 97\%~\cite{Kozlov:2014fqa}, depending on smearing size
and shear viscosity over entropy ratio $\eta/s$ used in simulations.
Consequently, one would expect that the integrated value of 
$\lambda_2$ to be smaller than 3\%.\footnote{
Actually, to obtain the value in \Eq{eq:lambda} the integration
with respect to $p_T$ should be done separately for numerator
($\int dp_T \bra|X_n(p_T)|^2 \ket dN/dp_T$)
and denominator
($\int dp_T2\kappa_n^2(p_T) \bra |\E_2|^2\ket dN/dp_T$), 
which leads to an even smaller ratio than 
$\int dp_T \lambda_2(p_T) dN/dp_T$.
} 
Similarly, the upper bound of
$\lambda_3$ can be taken as 5\%. 

In total, for the central events of p-Pb, correlation between
$V_2$ and $V_3$ is expected with a value \emph{greater} than 8\% 
if late stage collective expansion indeed dominates the evolution.

\section{Conclusions}
\label{sec:con}

In summary, I have studied correlations between harmonics in the 
small colliding systems: p-Pb, d-Au and $^3$He-Au. %In particular,
Emphasis is laid on the mixing between $V_2$ and $V_3$, in terms of
mixing between
ellipticity $\E_2$ and 
triangularity $\E_3$, by assuming linear eccentricity scaling. 
Significant anti-correlations are found between
$\E_2$ and $\E_3$ in p-Pb 
systems from Monte Carlo Glauber simulations, with $|\P_{23}^{\e\e}|$ of the order
of 10\% in the central collision events. %due to initial state fluctuations. 
When the background geometry exhibits an intrinsic ellipticity, such
as central d-Au collisions, or an intrinsic triangularity, such
as central $^3$He-Au collisions, the correlation is even
enhanced. Monte Carlo Glauber simulations present stronger anti-correlation of
$\P_{23}^{\e\e}$ in d-Au than that in $^3$He-Au. 
Similar, but stronger correlation patterns are found 
when the dipolar asymmetry $\E_1$ is involved.
The physical origin of initial state correlations is
analytically studied in the independent-source model, from which
fluctuations and intrinsic shape are recognized as the dominant
effects. 

Mixing between harmonic flow is basically considered equal to
the corresponding mixing of initial anisotropies, due to an
approximate linear eccentricity scaling. Effects beyond linear
eccentricity scaling are discussed by including noises in the 
linear flow response, from which a more realistic estimate of 
$\P_{23}^{vv}$ is made for the central p-Pb collisions.
Regarding measurements carried out by the CMS collaboration,
statistics in experiment is discussed in \App{app:sta}.
Mixing between $V_2$ and $V_3$ is thus proposed as
an accessible probe to detect the medium collectivity 
in experiments, if evolution
in p-Pb collisions is dominated by the medium collective
expansion. In d-Au and $^3$He-Au,
especially in the most central events, medium collectivity is 
supposed to lead to sizable $\P_{23}^{vv}$ as well, although
its patterns may be more involved due to other effects like non-linearities 
in the flow generation. However, 
geometries of p-Pb, d-Au and $^3$He-Au systems are 
expected to be revealed 
by comparing $\P_{23}^{vv}$ among these colliding systems.
%A systematic analysis with event-by-event hydrodynamics 
%simulations will be our future work.
In a similar manner, correlations involving $V_1$ are proposed as
probes as well, which are generated from initial correlations involving
$\E_1$.

\acknowledgements

I am grateful for many helpful discussions with Jean-Yves Ollitrualt
at different stages of this work.
This work is supported by
the European Research Council under
the Advanced Investigator Grant ERC-AD-267258. 

\appendix

\section{Statistical errors of $\P_{23}^{vv}$ measurements in p-Pb }
\label{app:sta}

In this appendix, I briefly estimate the statistical error on $\P_{23}^{vv}$
in the central p-Pb collisions at the LHC. The order of magnitude of 
the number of events that is needed for the observation of $\P_{23}^{vv}$ in
experiments can be inferred consequently.
The measurement of $\P_{23}^{vv}$ involves a five-particle correlation,
\be
\bra V_2^3(V_3^*)^2\ket
\ee
whose statistical error is determined by the total number of 
independent 5-plets which can be constructed. 
For a total number of events $N_{eve}$ and
multiplicity $M$ in each single event, statistical fluctuations
with respect to $\bra V_2^3(V_3^*)^2\ket$ are 
therefore of the order $1/(N_{eve}M^5)^{1/2}$.  
Statistical errors on $\P_{23}^{vv}$
from measurements can be accordingly estimated as
\be
\label{eq:error}
\delta \P_{23}^{vv}\sim 
\frac{1}{\sqrt{N_{eve}}}\left(\frac{1}{v_2 \sqrt{M}}\right)^3
\left(\frac{1}{v_3 \sqrt{M}}\right)^2\,.
\ee 
The rule of thumb of writing \Eq{eq:error} is that each $V_n$ contributes
a factor $1/(|V_n|\sqrt{M})$.
Similar analysis can be generalized to the measurements of other types of
correlations between harmonics.

Referring to the measurements of p-Pb carried out by the 
CMS collaboration~\cite{Chatrchyan:2013nka,Khachatryan:2015waa}, 
the resolution of $v_2$ and $v_3$
measurements for central collisions 
are approximately $(1/v_2\sqrt{M})\sim 1.25$ and 
$(1/v_3\sqrt{M})\sim 5$ respectively.
In total,
to observe $\P_{23}^{vv}$
with value at least 8\%, a total number
of one million events are roughly needed for each centrality bin.  
The LHC has already recorded an integrated luminosity 
of 35nb$^{-1}$(cf. \Ref{Khachatryan:2015waa}) in the 
p-Pb runs collected in 2012 and 2013, corresponding 
to a total events of the order of $10^{10}$.
Therefore there are over one million events in the 0.01\% most central bin, 
and the anti-correlation between $V_2$ and $V_3$ can be detected as 
a feasible probe for the test of medium collectivity.

%
%Considering the total number of 
%p-Pb collision events achieved at the LHC, 
%with an integrated luminosity of 35nb$^{-1}$(cf. \Ref{Khachatryan:2015waa}), 
%it is thus expected the anti-correlation between $V_2$ and $V_3$ 
%to be a feasible probe for the test of medium collectivity.

\section{$\P_{23}^{\varepsilon\varepsilon}$ from independent sources}
\label{app:indep}

We take complex notations for simplicity in the following derivations, 
so that the transverse coordinate is expressed as $z=r\exp{(i\phi_r)}$.
In the independent-source model, azimuthal eccentricities of a 
fluctuating initial state can be analytically written. 
Taking into account of the re-centering corrections, 
\begin{align}
\label{eps_def2}
%\E_1=&\varepsilon_1 e^{i\Phi_1}=-\frac{\{(z-\d_z)^2(\bar z-\d_{\bar z})\}}{\{r^3\}}\,,\nonumber\\
\E_2=&\varepsilon_2 e^{i2\Phi_2}=-\frac{\{(z-\d_z)^2\}}{\{r^2\}}\,,\nonumber\\
\E_3=&\varepsilon_3 e^{i3\Phi_3}=-\frac{\{(z-\d_z)^3\}}{\{r^3\}}\,.
\end{align}
\Eq{eps_def2} can be expanded with respect to fluctuations order by order, 
which in turn gives rise to an expansion 
in terms of $1/N$. Therefore we have,
\begin{align}
\label{eps_exp}
%\E_1=&-\frac{\{(z-\d_z)^2(\bar z-\d_{\bar z})\}}{\{r^3\}}\,,\nonumber\\
\E_2=&-\frac{[\bra z^2\ket + \d_{z^2} + \dz^2]}{\bra r^2\ket }+ O(\d^3)\,,\nonumber\\
\E_3=&-\frac{[\bra z^3\ket -3\bra z^2\ket\dz + \d_{z^3} - 3\dz\d_{z^2} ]}{\bra r^3\ket}+O(\d^3)\,.
\end{align}
Note that $\bra z^2\ket=-\bra r^2\ket {\bar \E}_2$ or 
$\bra z^3\ket = -\bra r^3\ket {\bar \E}_3$ in \Eq{eps_exp} 
vanishes when there is no corresponding intrinsic eccentricity from the
background geometry. For simplicity, contributions from fluctuations of denominators in \Eq{eps_def2}
are neglected, which does not affect the counting of power of fluctuations and is allowed 
quantitatively for p-Pb. 

In the independent source model, in addition to the two-point function in \Eq{eq:twop}
which is proportional to $1/N$,
multi-point function can be evaluated using Wick's theorem,
and found to be suppressed by
higher powers of $1/N$~\cite{Alver:2008zza}. 
Implicitly, one has
\be
\bra\delta^3\ket\sim\bra\delta^4\ket\sim O\left(\frac{1}{N^2}\right),\quad
\bra\delta^5\ket\sim\bra\delta^6\ket\sim O\left(\frac{1}{N^3}\right),\ldots
\,.
\ee
Pearson coefficients of initial state \Eq{pers_ini} can therefore be 
written order by order in powers of $1/N$.
For the p-Pb systems, because all terms proportional to $\bra z^2\ket$ or
$\bra z^3 \ket$ vanish, one has 
\be
\label{eq:p23pb-1}
\P_{23}^{\e\e}=-\frac{\bra \d_{z^2}^3\d_{{\bar z}^3}^2\ket + 3\bra\d_{z^2}^2 \d_z^2 \d_{{\bar z}^3}^2\ket
- 6\bra \d_{z^2}^3\d_{\bar z}\d_{{\bar z}^2}\d_{{\bar z}^3}\ket + O(\bra\d^7\ket)}
{\left[\bra \d_{z^2}^3\d_{{\bar z}^2}^3\ket + O(\bra \d^7\ket)\right]^{1/2}
\left[\bra \d_{z^3}^2 \d_{{\bar z}^3}^2\ket + O(\bra \d^5\ket)\right]^{1/2}}\sim
\frac{1}{N^{3/2}}\,.
\ee
The minus sign in \Eq{eq:p23pb-1} comes from convention used in the 
definition of anisotropies and $\P_{23}^{vv}$, which agrees with
the analysis given in \Sect{sec:dis}.   
Again, in evaluating multi-point functions, azimuthal symmetry
in p-Pb demands that only terms of the type $\bra r^{2n} \ket=\bra z^n{\bar z}^n\ket$
left, which leads to the result in \Eq{eq:p23ppb}. 
For $^3$He-Au, terms containing $\bra z^3\ket$ contribute. Besides, terms of the
type $\bra z^a{\bar z}^b\ket\ne 0$ also contribute if $|a-b|/3 = 0$, 1, etc., which
leads to 
\begin{align}
\label{eq:p23he-1}
\P_{23}^{\e\e}=&-\frac{\bra {\bar z}^3\ket^2\bra \d_{z^2}^3\ket +
3\bra {\bar z}^3\ket^2 \bra \d_{z^2}^2 \d_z^2\ket + 
2\bra {\bar z}^3\ket\bra \d_{{\bar z}^3} \d_{z^2}^3\ket+O(\bra\d^5\ket)}
{\left[\bra z^3 \ket^2 \bra {\bar z}^3\ket^2 + O(\bra \d^2\ket)\right]^{1/2}
\left[\bra \d_{z^3}^2 \d_{{\bar z}^3}^2\ket + O(\bra \d^5\ket)\right]^{1/2}}\nonumber\\
=&-\frac{\bra {\bar z}^3\ket^2 \bra z^6 \ket + 6\bra z^3 \ket^2\bra {\bar z}^3\ket^2 }
{\sqrt{6N} \bra z^3 \ket\bra {\bar z}^3 \ket \bra r^4 \ket^{3/2}} + O\left(\frac{1}{N^{3/2}}\right)\,.
\end{align}
For d-Au collisions, $\bra z^2\ket\ne 0$ and accordingly
all terms of the type $\bra z^a{\bar z}^b\ket\ne 0$ if $|a-b|$ is a even integer. 
One finds that the lowest order term is independent of $1/N$,
\begin{align}
\P_{23}^{\e\e}=-\frac{\bra z^2\ket^3\left(9\bra {\bar z}^2\ket^3 + \bra {\bar z}^6\ket
- 6\bra {\bar z^2}\ket \bra {\bar z^4}\ket\right)}
{\left[81\bra z^2\ket^2 \bra {\bar z}^2\ket^2\left(
2\bra z{\bar z}\ket^2 + \bra z^2\ket\bra {\bar z}^2\ket\right)
+2\bra z^3{\bar z}^3\ket^2  +\ldots\right]^{1/2}}+
O\left(\frac{1}{\sqrt{N}}\right)
%+ \bra z^6\ket\bra{\bar z}^6 \ket
%+\bra z^2\ket\bra {\bar z}^2\ket\left(
%\bra z{\bar z}\ket\bra z^3{\bar z}^3\ket+\bra z{\bar z}^3\ket\bra{\bar z}z^3 \ket
%\bra z^4\ket \bra {\bar z}^4\ket\right)\right]^{1/2}}
\end{align}

\bibliography{refsbib}

\end{document}